\documentclass[twocolumn,twoside,slac]{revtex4}
\usepackage{graphicx}
\usepackage{fancyhdr}
\begin{document}

\title{Sensitivity of searches for new signals and its optimization}

%

\author{ Giovanni Punzi}
\affiliation{ Scuola Normale
Superiore and INFN, Pisa, Italy}

\begin{abstract}
A frequentist definition of sensitivity of a search for new phenomena is discussed, that
has several useful properties. It is based on completely standard concepts, is generally
applicable, and has a very clear interpretation. It is particularly suitable for optimization,
being independent of a-priori expectations about the presence of a signal, thus allowing
the determination of a single set of cuts that is optimal both for setting limits and for 
making a discovery. Simple approximate formulas are given for the common problem of Poisson 
counts with background.

\end{abstract}

\maketitle

\thispagestyle{fancy}


\section{\label{sec:Intro}Introduction} 

The question of the sensitivity of a search for new phenomena is a very common one. The need may arise either by the wish to predict the outcome of an experiment and compare several possible experiments or different configurations of the same experiment. Several different ways have been used to quantify the sensitivity of a search, which makes it sometimes difficult to compare them. In particular, two different sensitivity figures are often quoted, one that is relative to the potential for actually making a discovery, and another to characterize how strong a constraint is imposed on the unknown phenomena if no evidence is found for a deviation from the standard theory. This situation makes it difficult to optimize the design of an experiment, because it is not clear what should be maximized. I describe here a definition of sensitivity which is unique and well-defined for any experiment. This is based on purely frequentist ideas, which avoids the issue of the choice of an a-priori distribution for a new and unknown phenomena.

\section{Statement of the problem}

The problem of searches for new phenomena can be stated formally in classical statistics as one of ``Hypothesis testing". We have a ``default hypothesis" $H_0$, that is our current best theory, and as a result of the experiment we wish to either confirm or disprove the theory $H_0$, in favor of an alternative theory $H_m$, where $m$ indicates the free parameters of the new theory (mass or set of masses of new particles, coupling constants, production cross sections, etc.). The experiment consists of measuring the value of a set of observables $X$ (possibly a large number) whose distribution depends on the true state of nature  being $H_0$ or $H_m$. In a simple counting experiment, the observable $X$ is the number of observed counts, and hypothesis $H_0$ is defined as the distribution of $X$ being a Poisson with the mean equal to the number of expected background events $B$. Hypothesis $H_m$ is that the distribution is instead a Poisson with a larger mean $B+S_m$, where $S_m$ is the expected contribution of the ``new signal", which is a function of the unknown free parameters of the new theory, $m$.
A test of $H_0$ is specified by defining the set of values of $X$ that will make us decide that $H_0$ must be rejected (``critical region"); the {\em significance level} of the test, indicated by $\alpha$, is the probability of  rejecting $H_0$ when it is indeed true; that is to say, $\alpha$ is the probability for $X$ to fall within the critical region, calculated under the assumption that $H_0$ is true. There are many possible choices of the critical region, therefore many possible different tests at the given significance level $\alpha$, and we will not be concerned here with the way the choice is made; all of the present discussion  is independent of the way the test was chosen.

What about the value of  $\alpha$ ? This is a ``small number", common practice for really new physics discovery being to require $\alpha$ to correspond to the $5\sigma$ single tail of a gaussian distribution.

The other element to be considered in a test is the probability that a discovery is made. The classical way to express this is by the {\em power function} $1-\beta(m)$, that is, the probability that $X$ will fall in the critical region (=the probability that a discovery will be claimed) assuming $H_m$ is true, as a function of the parameters $m$. 
It is clearly desirable to have the greatest possible power. However, it is well known that only in very few special problems it is possible to maximize the power simultaneously for every $m$. For this reason, trying to optimize the power is subject to a judgement about what values of the parameters are more important; in the next section we will show how to solve the issue by attacking the problem from a different angle.

After a measurement is performed, if no discovery is made the experimenter will usually produce an additional piece of information: a confidence region for the unknown parameters $m$. This part is in principle completely independent from the ``testing" part, and interesting issues arise when one tries to make sure the two kinds of information are coherent. For instance, limits are often desired at a confidence level lower than the level of significance required for claiming a discovery; this can lead easily to situations where no discovery is claimed, and yet limits are quoted that do not include the $H_0$ hypothesis. For the purpose of the present discussion we don't need to deal with such difficult issues and we will make only minimal assumptions about the relationship between the test and the algorithm adopted for setting limits. We will just assume that the confidence band for $m$ be built in such a way to exclude, whenever possible, all values of $X$ falling within the acceptance region for $H_0$; (this can be done for every $m$ such that $1-\beta(m)>CL$, where CL is the desired Confidence Level).  This is quite natural, and usually happens spontaneously, because it makes for tighter confidence regions when no discovery is made, at no expense.

If a discovery is indeed made, the most interesting piece of information in the result will be the discovery itself, and maybe an estimate of the parameters $m$, so we will not be concerned with limit setting in case of discovery, only with the probability that it happens.

\section{\label{sec:sensitivity}Definition of Sensitivity of a search experiment}

Many definitions of sensitivity for a search have to do with either the "average limit" produced if $H_0$ is true (defined in various ways), or with the significance of an observed signal, assuming the observation is exactly equal to the expected value in presence of a signal at $m$.

We suggest to characterize the sensitivity of an experiment in the following way. Correct statistical practice requires to decide before the experiment the values of $\alpha$ and CL, so we assume their values are given. Then one can proceed by quoting the region of the parameters $m$ for which {\em the power of the chosen test is greater or equal to the Confidence Level chosen for the limits in case there is no discovery}:
\begin{equation}\label{eq:sens}
1-\beta_\alpha(m)>CL
\end{equation}

This region of $m$ can be thought of as a region of parameters to which the experiment is ``sufficiently sensitive". While it is always possible to provide additional information by plotting contours of constant power in the $m$ space for values different from the CL, the specific region defined by eq.~(\ref{eq:sens}) is particularly informative because it has a very simple and clear-cut interpretation. In fact, it is easy to verify that the following two statements hold simultaneously:

\begin{itemize}
\item If the true value of $m$ satisfies~(\ref{eq:sens}), then there is a probability at least $CL$ that performing the experiment will lead to discovery (with the chosen significance $\alpha$).

\item If performing the experiment does not lead to discovery, the resulting limits will exclude (at least) the entire region defined by~(\ref{eq:sens}), at the chosen CL. (N.B. this relies on the minimal assumption of a ``reasonable algorithm" for setting limits made in previous section, and holds independently of the true value of $m$.)

\end{itemize}

In short, eq.~(\ref{eq:sens}) defines the region in the parameter space for which the experiment will {\em certainly} give an answer: that region will be excluded, or a discovery will be claimed, with no possible in-between. This double discovery/exclusion interpretation suggests that it deserves to be named {\em sensitivity region} for the experiment and to be quoted as the single most useful information to characterize its potential and optimize it. Note explicitly that there is no possibility for an experimental fluctuation to jeopardize the result; it is possible for a fluctuation to increase the region of exclusion, but not to diminish it. In particular, if the parameter region covers the whole range of physically interesting values for $m$, the experiment can very well been said to be conclusive. This {\em sensitivity region} appears to be a more useful information than others commonly quoted,  that have a more vague meaning, like:

\begin{itemize}
\item the ``average" excluded region, {\em if} $H_0$ is true (tells you nothing certain about the actual limits that will be quoted; tells you nothing about what will happen if the signal exists but it is small)
\item an "average number of sigmas", for given values of $m$, or the number of sigmas you would get in case exactly the expected number of signal events is observed (tells you nothing about the limits in case there is no observation; tells you little about how likely it is that a signal will actually be observed, due to the effect of statistical fluctuations)
\end{itemize}

Comparison between two experiments, or experimental settings, should be made on the basis of whether one sensitivity region includes the other. It is still possible for two experiments to be non-comparable, by having none of the two region completely include the other; in that case, the issue of which is preferable cannot be resolved on a statistical basis, but it is a question of strategy. If the sensitivity regions are very different, the actual conclusion is that the two experiments are somehow `complementary', probing different regions of the parameters space.

There are a few other arguments in favor of quoting this quantity to characterize the sensitivity of an experiment:

\begin{itemize}

\item The definition is independent of the choice of metric (in both observable and parameter space).

\item It does not require a choice of priors

\item It is straightforward (and meaningful) to apply even in complex situations. For instance:
	\begin{itemize}
		\item 1-D problems with a ``non monotonic" structure. Example: search for a CP violation effect, where one measures the sine of an angle, with the range $[-1,1]$. In this case $H_0$ is in the middle, and it makes no sense to quote ``average upper limit".
		\item multidimensional parameter problems. Examples of this kind are neutrino oscillation searches, where the space is 2-D. Even more complex examples are found in CP-violation measurements in neutral B mesons oscillations, where both a direct and a mixed component are possible; in this case the allowed region for the parameters is circle of unit radius, $H_0$ being at the center, and it is impossible to use concepts like ``average upper limit", or even ``median of the limit".
	\end{itemize}

\item It is independent of the expectations for a signal to be present, thus allowing an unbiased optimization.

\item It allows you to optimize what you really want for a search, without being distracted by other elements. For instance, if one had to concentrate on getting the maximum possible power (e.g. by looking at its average  it over a chosen region), one can easily be fooled into preferring an experiment that has a very high power in a region where the power is pretty high anywyay, over one that has a more even distribution of power, that is actually much more likely to provide useful information, since in a discovery measurement the power counts the most where it is ``intermediate". Considering the region rather than power in itself takes this into account.

\end{itemize}

\section{\label{sec:examples}Optimization of a counting experiment}

We will now apply the ideas discussed in the previous section to the very common problem of a counting experiment in presence of background. In this case, we have the discrete observable $n$, the number of events observed, which is Poisson-distributed with a mean determined by $B$, the expected number of background events (supposed known), and the possible contribution of signal events $S_m$:
\begin{eqnarray}
p(n|H_0) = &e^{-B} B^n/n!\\
p(n|H_m) = &e^{-B-S_m} (B+S_m)^n/n! 
\end{eqnarray}
For this problem, the only sensible definition of a critical region for the presence of non-zero signal $S_m$ takes the form of a condition like $$n>n_{min}$$

\begin{figure}[h]
\includegraphics[width=\linewidth,height=1.5in]{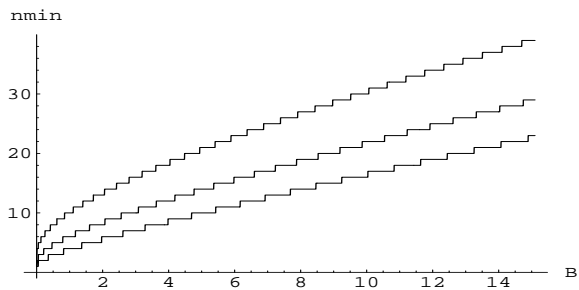}%
\caption{\label{fig:cut} Minimum number of observed events needed to claim discovery with 95\%, $3\sigma$, $5\sigma$ significance, vs expected background.}%
\end{figure}

Therefore, the test is completely defined once the desired significance level $\alpha$ is chosen. Figure~\ref{fig:cut} shows the value of $n_{min}$ as a function of $B$, for given values of $\alpha$, obtained by numerical calculation of sums of Poisson probabilities.

Having completely defined the test, we can now evaluate its power as a function of $m$, and determine the set of values for $m$ such that eq.~(\ref{eq:sens}) holds. Since the power of a test of the form $n>n_{min}$ grows monotonically with $S_m$, it is easy to see that eq.~(\ref{eq:sens}) leads to simple inequalities of the form:
$$S_m>S_{min}$$

\begin{figure}[h]
\includegraphics[width=\linewidth,height=1.5in]{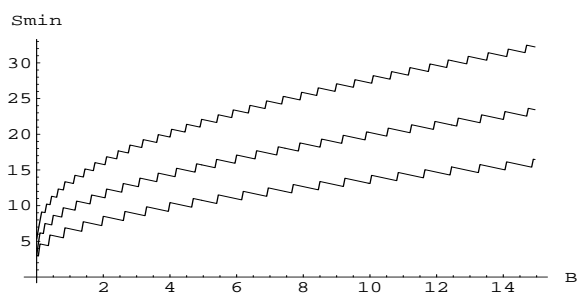}%
\caption{\label{fig:Smin} The lower limit of the sensitivity region $S_{min}$, for a search experiment with (significance, CL) respectively of (95\%,95\%), ($3\sigma$,95\%), ($5\sigma$,90\%).}%
\end{figure}

Therefore, all is needed to completely characterize the solution of our problem is the value of $S_{min}$, that is in general a function of $\alpha , \beta$, and $B$. Plots of $S_{min}$ from numerical calculation are shown in Fig.~\ref{fig:Smin}. 

Tables of this kind of data can in principle be used to compare different experimental settings, by determining for each of them the set of values of  $m$ such that $S_m>S_{min}$, and choosing the one with the largest set. 
However, it is much easier to perform such optimizations tasks with the help of an analytic parametrization. For the purpose of optimization, an approximation of the exact result is usually sufficient; in particular, there is no need to account for the discretization effects. 

A simple parametrization of our result can be obtained by means of Gaussian approximation of the Poisson. It is easy to see that in this approximation, condition~(\ref{eq:sens}) translates into the following equation for $S_{min}$:
\begin{equation}\label{eq:Gaussapprox}
S_{min} = a \sqrt{B} + b \sqrt{B + S_{min}}
\end{equation}

where $a$ and $b$ are the number of sigmas corresponding to one-sided Gaussian tests at significance $\alpha$ and $\beta$ respectively.

Solving eq.~(\ref{eq:Gaussapprox}) for $S_{min}$ yields the solution:

\begin{equation}
S_{min} = \frac{b^2}{2}+ a\,{\sqrt{B}} + 
  \frac{b}{2}\,{\sqrt{b^2 + 4\,a\,{\sqrt{B}} + 4\,B}}
\end{equation}


This expression holds for one specific set of data selection criteria. Now consider the common situation where one has to decide on the set of cuts to be used in the analysis. This means that both the background $B$ and the number of expected signal events $S_m$ will depend on the cuts (let's indicate the whole set of cuts with the symbol $t$). 
In a completely general case, in order to decide which set of cuts $t$ is best, one needs to determine for every $t$ the set of values $\tilde{m}$ to which the experiment is sensitive, by solving for $\tilde{m}$ the inequality:
$$ S_{\tilde{m}}(t) \ge \frac{b^2}{2}+ a\,{\sqrt{B(t)}} + 
  \frac{b}{2}\,{\sqrt{b^2 + 4\,a\,{\sqrt{B(t)}} + 4\,B(t)}}$$
  
and then choose the cuts $t$ yielding the most extended region.
The situation is much simpler when the efficiency $\epsilon$ of the chosen cuts on the signal is indipendent of $m$, that is when one can write:
$$S_m(t) = \epsilon(t)\cdot L \cdot \sigma_m$$

where $L$ is the integrated luminosity and $\sigma_m$ is the cross section of the process being searched for.

In this case one can simply invert the above equation to write down the minimum ``detectable" (according to our criteria) cross section:
$$ \sigma_{min} = \frac{ \frac{b^2}{2}+ a\,{\sqrt{B(t)}} + 
  \frac{b}{2}\,{\sqrt{b^2 + 4\,a\,{\sqrt{B(t)}} + 4\,B(t)}}}{ \epsilon(t)\cdot L}$$

Obviously, the maximum sensitivity is attained when $\sigma_{min}$ is smallest, that is when the quantity:
\begin{equation} \label{eq:completemax}
 \frac{ \epsilon(t)}{ b^2+ 2\,a\,{\sqrt{B(t)}} + 
  b\,{\sqrt{b^2 + 4\,a\,{\sqrt{B(t)}} + 4\,B(t)}}}
\end{equation}

reaches its maximum. Note explicitly that, in the given assumption of the efficiency being independent of $m$, the optimal choice of cuts {\em does not depend} on the assumed cross section for the new process $\sigma_m$. This is a very useful feature, since this parameter is often unknown, and it is a direct consequence of the chosen approach, that focuses on maximizing the power where it is really necessary, that is at the threshold of visibility.
Expression~(\ref{eq:completemax}) becomes even simpler when the choice $b=a$ is made:
\begin{equation} \label{eq:simplemax}
 \frac{ \epsilon(t)}{ a/2 + \sqrt{B(t)}}
\end{equation}

This simple expression is adequate in most problems of search optimization; also, it is  readily compared with some ``significance-like" expressions that are commonly used for optimization purposes:

\begin{itemize}
\item[a)] $\frac{S}{\sqrt{B}}$
\item[b)] $\frac{S}{\sqrt{B+S}}$
\end{itemize}

Note that expression b) cannot be maximized without knowing explicitly the cross section for the searched signal. Also, it does not quite represent what one wants to maximize for a search, being more directly related to the relative uncertainty in the measurement of the yield of a new process, if found, than to significance. Expression a), being linear in $S$, shares with expression (\ref{eq:simplemax}) the good property of being independent of the cross section of the new process, but it has the important problem of breaking down at small values of $B$. Imposing maximization of a) may push the experiment efficiency down to very small values. In order to see the failure of expression a), it is sufficient to consider, for instance, that it will prefer an expectation of $0.1$ signal events with a background of $10^{-5}$ over a situation with $10$ signal events expected and a background of $1$ event. 

It should be apparent that expression~(\ref{eq:simplemax}) (or its slightly more sophisticated form~(\ref{eq:completemax})), compared with ``significances"  a) and b), is not only better motivated, but also unambiguously preferable from a practical viewpoint.

The features of the discussed formulas are more easily seen by plotting the factor $1/S_{min}$ from the exact calculation (that is proportional to the quantity that needs to be maximized, as in eq.~(\ref{eq:completemax})) together with the two significance--like expressions discussed above: they all behave as $1/\sqrt{B}$ at large $B$, and it is therefore possible to normalize them to converge as $B\rightarrow \infty$. Expression b) is not simply proportional to $S$, so we had to make a choice and we put $\frac{S}{\sqrt{B+S}} = a$ , in agreement with the spirit of our current approach of focusing on the point where significance is at the threshold, and solved for $1/S$ to obtain a function of $B$ only.

\begin{figure}[h]
\includegraphics[width=\linewidth]{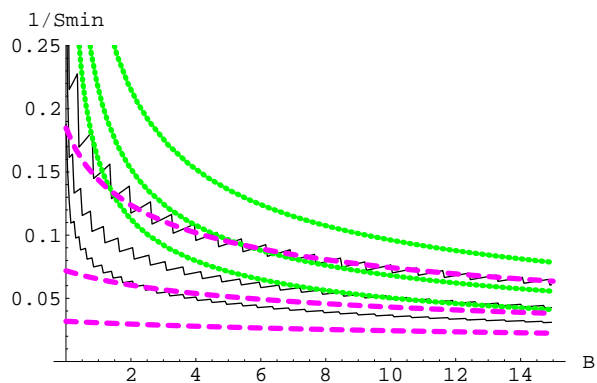}%
\caption{\label{fig:signif} Comparison of $1/S_{min}$ with the corresponding sensitivity factor given by $S/\sqrt{B}$ (dotted) and $S/\sqrt{S+B}$ (dashed), for a search experiment with (significance, CL) respectively of (95\%,95\%), ($3\sigma$,95\%), ($5\sigma$,90\%)}
\end{figure}

The comparison is shown in Fig.~\ref{fig:signif}, where it appears that our suggested solution lies between a) and b), where a) largely overestimates the ``sensitivity" at low backgrounds, as expected, and conversely b) underestimates it, expecially for high significance settings.

\begin{figure}[h]\includegraphics[width=\linewidth]{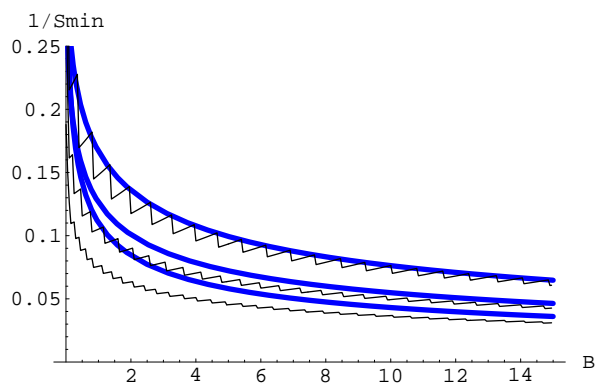}%
\caption{\label{fig:sens1} Gaussian approximation of the ``Sensitivity factor" $1/S_{min}$ (eq.~(\ref{eq:completemax})) for a search experiment with (significance, CL) respectively of (95\%,95\%), ($3\sigma$,95\%), ($5\sigma$,90\%)}%
\end{figure}

\begin{figure}[h]
\includegraphics[width=\linewidth]{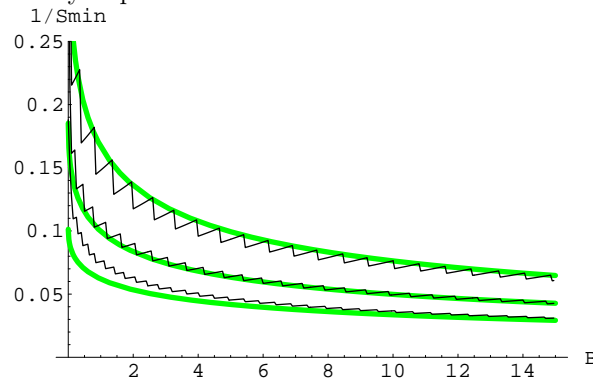}%
\caption{\label{fig:sens0} Gaussian approximation of $1/S_{min}$ in the $b\approx a$ approximation (eq.~(\ref{eq:completemax})), for a search experiment with (significance, CL) respectively of (95\%,95\%), ($3\sigma$,95\%), ($5\sigma$,90\%). Curves are normalized to the asymptotic limit.}
\end{figure}

The Gaussian approximation to the exact solution is shown instead in fig.~\ref{fig:sens1}, and its special case for $b\approx a $ in fig.~\ref{fig:sens0}. 


It can be seen that the approximate formulas work well at moderate values of $a$ and $b$, but become less accurate when high significance/CL are desired, due to the larger deviations from Gaussian behavior that occur in the Poisson far tails. However, the Gaussian approximation can easily be improved, without losing the good features of the solutions. For instance, it is possible to obtain a more accurate expression by accounting for differences between Gaussian and Poisson tail integrals at the next order in $a$ and $b$, simply by performing an empirical fit.
This results in the following improved expression for $S_{min}$:
\begin{equation}\label{eq:improvedGauss}
S_{min} = \frac{a^2}{8} + \frac{9\,b^2}{13} + a\,{\sqrt{B}} + 
  \frac{b}{2}\,\sqrt{b^2 + 4\,a\,{\sqrt{B}} + 4\,B}
            \end{equation}

\begin{figure}[h]
\includegraphics[width=\linewidth]{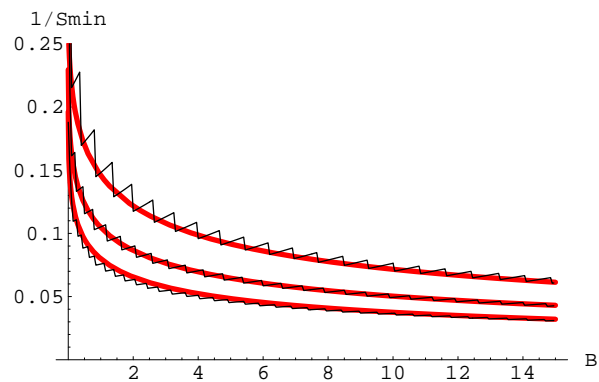}%
\caption{\label{fig:sens2} Improved Gaussian approximation of the ``Sensitivity factor" $1/S_{min}$ (eq.~(\ref{eq:improvedGauss}) for a search experiment with (significance, CL) respectively of (95\%,95\%), ($3\sigma$,95\%), ($5\sigma$,90\%)}
\end{figure}

Fig.~\ref{fig:sens2} shows this slightly modified expression to be considerably accurate even at high significances, which makes it suitable also for searches of ``really new" effects, where a significance level of $5\sigma$ is a customary requirements.

\begin{acknowledgments}

The author wishes to thank Louis Lyons for many helpful comments.

\end{acknowledgments}




\end{document}